\documentclass[aps,preprint,nofootinbib,%
superscriptaddress,amsmath,amssymb,showpacs,floatfix]{revtex4}
\usepackage{epsfig}
\usepackage{bm}
\usepackage{epsfig,graphics,graphicx}
\begin{document}
%
\preprint{}
\newcommand{\U}{\mathrm{U}}
\newcommand{\Th}{\mathrm{Th}}
\newcommand{\K}{\mathrm{K}}
\title{How much Uranium is in the Earth? \\Predictions for geo-neutrinos at KamLAND}
\author{Gianni Fiorentini}
\email{fiorenti@fe.infn.it} \affiliation{Dipartimento di Fisica,
Universit\`a di Ferrara, I-44100 Ferrara, Italy}
\affiliation{Istituto Nazionale di Fisica Nucleare, Sezione di
Ferrara, I-44100 Ferrara, Italy}
\author{Marcello Lissia}
\email{marcello.lissia@ca.infn.it} \affiliation{Istituto Nazionale
di Fisica Nucleare, Sezione di Cagliari,
             I-09042 Monserrato, Italy}
\affiliation{Dipartimento di Fisica, Universit\`a di Cagliari,
             I-09042 Monserrato, Italy}
\author{Fabio Mantovani}
\email{mantovani@fe.infn.it} \affiliation{Dipartimento di Scienze
della Terra, Universit\`a di Siena, I-53100 Siena, Italy}
\affiliation{Centro di GeoTecnologie CGT,I-52027 San Giovanni
Valdarno, Italy} \affiliation{Istituto Nazionale di Fisica
Nucleare, Sezione di Ferrara, I-44100 Ferrara, Italy}
\author{Riccardo Vannucci}
\email{vannucci@crystal.unipv.it} \affiliation{ Dipartimento di
Scienze della Terra, Universit\`a di Pavia, I-53100 Siena, Italy}
\affiliation{Centro di GeoTecnologie CGT,I-27100 Pavia, Italy}
%

\date{January 4, 2005; revised July 19, 2005}
\begin{abstract}

Geo-neutrino detection can  determine the amount of  long-lived
radioactive elements within our planet, thus  providing a direct
test of the Bulk Silicate Earth (BSE) model and fixing  the
radiogenic contribution to the terrestrial heat. We present a
prediction for the geo-neutrino signal at KamLAND as a function of
the Uranium mass in the Earth. The prediction is based on global
mass balance, supplemented by a detailed geochemical and
geophysical study of the region near the detector.  The prediction
is weakly dependent  on mantle modeling. If BSE is correct,
Uranium geo-neutrinos  will produce between 25 and 35 events per
year and $10^{32}$  protons at Kamioka.

\end{abstract}

\maketitle
\section{\label{sec:intro}Introduction}

The deepest hole that has ever been dug is about 12~Km deep, a
mere dent in planetary terms.  Geochemists analyze samples from
the Earth's crust and from  the top of the  mantle, whereas
seismology can reconstruct the density profile  throughout the
whole Earth but not its composition. In this respect,  our planet
is mainly unexplored. Geo-neutrinos --- the antineutrinos from the
progenies of U, Th and $^{40}$K decays in the Earth --- bring to
the surface information from the whole planet, concerning its
content of radioactive elements. Their detection can shed light on
the sources of the terrestrial heat flow, on the present
composition and on the origins of the Earth.

Geo-neutrino properties, summarized in
Table~\ref{table:nuproperties}, deserve a few comments:

(i) geo-neutrinos originating  from different elements can be
distinguished due to their different energy spectra, e.g.,
geo-neutrinos with  energy $E>2.25$~MeV are produced only from the
Uranium chain;

(ii) geo-neutrinos from U and Th (not those from $^{40}$K) are
above threshold for the classical anti-neutrino detection
reaction, the inverse beta on free protons:

\begin{equation}\label{eq:inversebeta}
\bar{\nu} + p \to e^{+} + n - 1.8\mathrm{\ MeV} \quad ;
\end{equation}

(iii) anti-neutrinos from the Earth are not obscured by solar
neutrinos, which cannot yield reaction~(\ref{eq:inversebeta}).

\begin{table*}[htb]
\caption{The main properties of geo-neutrinos:  $\epsilon_{H}$
 is the heat production
rate per unit mass and natural isotopic composition;
$\epsilon_{\bar{\nu}}$ is the antineutrino production rate (number
of antineutrinos per unit time) per unit mass (Contribution of
$^{235}$U is neglected due to the small, 0.7\%, natural
abundance).} \label{table:nuproperties}
\newcommand{\dg}{\hphantom{$0$}}
\newcommand{\cc}[1]{\multicolumn{1}{c}{#1}}
\renewcommand{\arraystretch}{1.2} 
\begin{tabular}{llllll}
\hline Decay           & \cc{$Q$} & \cc{$\tau_{1/2}$}
& \cc{$E_{\mathrm{max}}$} & \cc{$\epsilon_{H}$} & \cc{$\epsilon_{\bar{\nu}}$}\\
      &  \cc{[MeV]} & \cc{[$10^9$~yr]} & \cc{[MeV]} &
      \cc{[W/Kg]} & \cc{[kg$^{-1}$s$^{-1}$]} \\
\hline $^{238}\mathrm{U\phantom{h}} \to {}^{206}\mathrm{Pb} + 8\,
{}^{4}\mathrm{He} + 6 e + 6 \bar{\nu}$
&   51.7 & \dg4.47& 3.26 & $0.95\times 10^{-4}$ & $7.41\times 10^{7}$ \\
$^{232}\mathrm{Th} \to {}^{208}\mathrm{Pb} + 6\, {}^{4}\mathrm{He}
+ 4 e + 4 \bar{\nu}$
&   42.7 &   14.0 & 2.25 & $0.27\times 10^{-4}$ & $1.63\times 10^{7}$ \\
$^{\phantom{0}40}\mathrm{K\phantom{h}} \to
{}^{\phantom{0}40}\mathrm{Ca} + e + \bar{\nu}$
& \dg1.32& \dg1.28& 1.31 & $0.36\times 10^{-8}$ & $2.69\times 10^{4}$ \\
\hline
\end{tabular}\\[2pt]
\end{table*}

Geo-neutrinos  were introduced by Eder~\cite{Eder} in the sixties
and Marx~\cite{Marx} soon realized their relevance. In the
eighties Krauss et al. discussed their potential as probes of the
Earth's interior in an extensive publication~\cite{Krauss:zn}. In
the nineties the first paper on a geophysical journal was
published by Kobayashi et al.~\cite{Kobayashi}.  In 1998, Raghavan
et al.~\cite{Raghavan:1997gw} and Rotschild et
al.~\cite{Rothschild:1997dd} pointed out the potential of KamLAND
and Borexino for  geo-neutrino detection.

In the last three years more papers appeared than in the previous
decades:  in a series of papers Fiorentini et
al.~\cite{Fiorentini:2002bp,Fiorentini:2003ww,Mantovani:2003yd,Fiorentini:2004xk,Fiorentini:2004rj}
discussed the role of geo-neutrinos for determining the radiogenic
contribution to the terrestrial heat flow and for discriminating
among different models of Earth's composition and origin.  A
reference model for geo-neutrino production, based on  a
compositional map of the Earth's crust and on geochemical modeling
of  the mantle, was presented in~\cite{Mantovani:2003yd}. At the
end of 2002, the analysis of the first data release of
KamLAND~\cite{Eguchi:2002dm} (equivalent  to an exposure of
$0.11\times 10^{32}$ proton~yr and 100\%  efficiency) reported 4
events from Uranium and 5 from Thorium out of a total of 32 counts
in the geo-neutrino energy region ($E_{\mathrm{vis}}<2.6$~MeV),
after subtracting 20 reactor events and  3 background  counts.
Statistical fluctuations imply an error of,  at least,  5.7
counts. Indeed, this first indication of geo-neutrinos stimulated
several
investigations~\cite{Nunokawa:2003dd,Mitsui:2003fm,Miramonti:2003hw,%
Domogatski:2004gs,McKeown:2004yq,Fields:2004tf,Rusov:2003sx,Fogli:2004vb}.

In a few years  KamLAND should  provide definite evidence of
geo-neutrino signal, after accumulating a much larger statistics
and reducing background. In the meanwhile other projects for
geo-neutrino detection are being developed. Borexino at Gran
Sasso, which is expected to take data in a few years, will benefit
from the absence of nearby reactors. Domogatski et
al.~\cite{Domogatski:2004zz} are proposing a 1 Kton scintillator
detector in Baksan, again very far from nuclear reactors. A group
at the Sudbury Neutrino Observatory in Canada is studying the
possibility of using liquid scintillator after the physics program
with heavy water is completed. The LENA proposal envisages a 30
Kton liquid scintillator detector at the Center for Underground
Physics in the Pyhas{\"a}lmi mine (Finland). Due to the huge mass,
it should collect several hundreds of events per year. In
conclusion, one can expect that within 10 years the geo-neutrino
signal from Uranium and Thorium will be measured  at a few points
on the globe.

In this paper  we shall concentrate  on geo-neutrinos from
Uranium, which are closer to   experimental detection, and on the
predictions for Kamioka, the site hosting the only detector which
is presently operational. \emph{Our goal is to understand which
information on the total amount of Uranium in the Earth can be
extracted from geo-neutrino measurements}.

As briefly discussed in the next section, the Uranium mass in the
Earth is estimated on the grounds of cosmo-chemical arguments,
based on the compositional similarity between Earth and
carbonaceous chondrites.   Measurements of samples from the
Earth's crust imply that the crust contains about one half of this
global estimate, whereas  the mantle --- which should contain the
rest --- is practically unexplored in this respect. A direct
determination of the Uranium mass in the globe is clearly an
important test of the origins of the Earth. Furthermore, such a
determination  will also fix  the radiogenic contribution to the
terrestrial heat flow, which is a presently debated issue, see,
e.g., Ref.~\cite{Anderson}. Early estimates of the geo-neutrino
signal and their connection with the global Uranium content are
also reviewed at the end of Section~\ref{sec:inTheEarth}.

The geo-neutrino signal depends on the total Uranium mass $m$ in
the Earth and on the geochemical and geophysical properties of the
region around the detector~\cite{Fiorentini:2002bp}. For KamLAND,
we estimated~\cite{Mantovani:2003yd} that about one half of the
signal is originated within 200~km  from the detector. This
region, although containing a globally negligible  amount of
uranium, produces a large contribution to the signal as a
consequence of its proximity to the detector. This contribution
has to be determined on the grounds of a detailed geochemical and
geophysical study of the region, if one wants to extract from the
total signal the remaining part which carries the relevant
information on $m$.   The study of the region around Kamioka,
including the possible effects of the subducting plates below the
Japan Arc and a discussion of the contribution from  of the Japan
Sea, is presented in Section~\ref{sec:nearTheDetector}.

The contribution from the rest of the world, discussed in
Section~\ref{sec:restOfTheWorld},  depends on  the total mass of
Uranium as well as on its distribution inside the Earth, since the
closer is the source to the detector the larger is its
contribution to the signal. For each value of $m$, we  shall
construct the distributions of Uranium abundances which provide
the maximal and minimal signals, under the condition that they are
consistent with geochemical and geophysical information on the
globe. For the Earth's crust we shall use a $2^\circ \times
2^\circ$ map~\cite{mappa} which distinguishes several crustal
layers and to each layer we shall assign minimal/maximal values
for the Uranium mass abundances. According to geochemists,  the
rest of Uranium should be found in the mantle. Observational data
for this reservoir are very poor, however  it is generally
believed that Uranium abundance increases with depth. The
assumption \emph{ that abundance is spherically symmetrical and
non decreasing with depth  will be enough  to provide rather tight
bounds on the mantle contribution to the geo-neutrino signal}.
This will be further combined with  the results for the crust.

\begin{figure}[htbp]
\includegraphics[width=0.7\textwidth,angle=90]{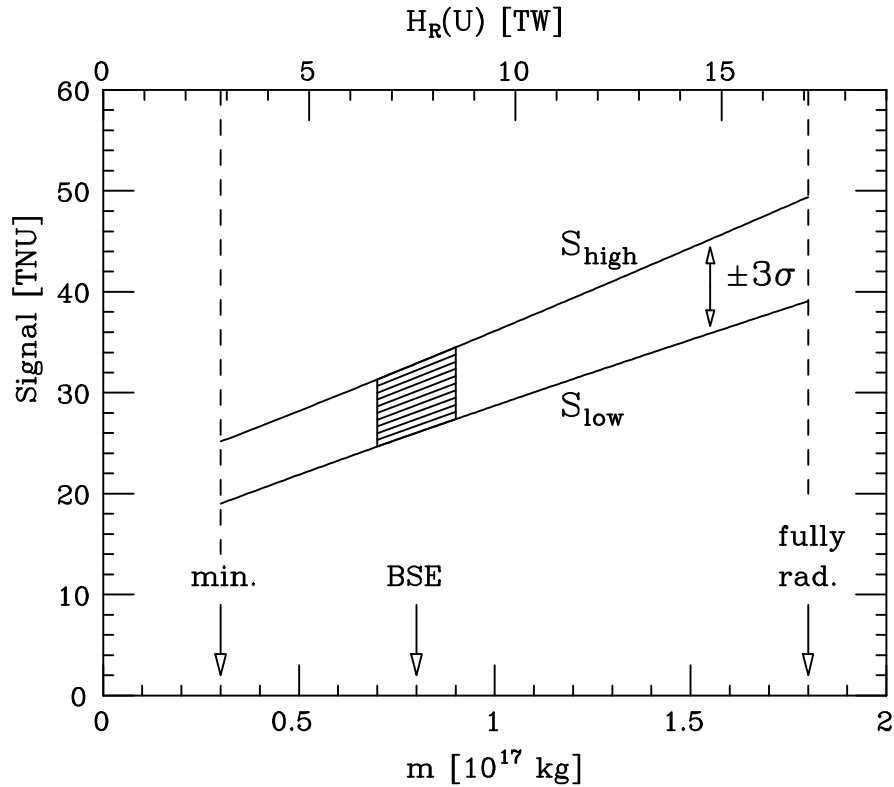}
\caption{The predicted signal from Uranium geo-neutrinos at
KamLAND.} \label{fig:Kamlandsignalmass}
\end{figure}

We shall put together the pieces of the above analysis in
Section~\ref{sec:signalVsMass}, where we present our main result
in Fig.~\ref{fig:Kamlandsignalmass}: a  narrow band describes the
predicted signal as a function of Earth's total Uranium mass. We
remark that the extremes of the band correspond to the whole range
of uncertainty, which is estimated according to the following
criteria: (i) for statistical errors we consider a $\pm 3\sigma$
interval; (ii) for systematic uncertainties of geochemical and
geophysical parameters we determine an interval such as  to cover
all modern estimates which we found in the literature; (iii)
independent errors are combined in quadrature.

On the grounds of Fig.~\ref{fig:Kamlandsignalmass} we discuss how
the geo-neutrino signal can provide a direct test of a fundamental
paradigm on the origins and composition of our planet.

\section{\label{sec:inTheEarth}U, Th and K in the Earth: how much and where?}

Earth's global composition is generally estimated from that of
chondritic meteorites by using geochemical arguments which account
for losses  and fractionation during planet formation. Along these
lines the Bulk Silicate Earth (BSE) model is built, which
describes the ``primitive mantle'', i.e., the outer portion of the
Earth after  core separation and before  the differentiation
between crust and mantle. The model is believed to describe the
present crust plus mantle system. Since lithophile elements should
be absent in the core~\footnote{One needs to be careful, since the
definition of an element's behavior, i.e., lithophile or not,
depends on the surrounding system; there exist models of the
Earth's core suggesting it as repository for radioactive
elements.}, the BSE provides the total amounts of U, Th and K in
the Earth, estimates from different authors being concordant
within 10-15\%~\cite{Mcdonough:2003}. From the estimated masses,
the present radiogenic heat production rate $H_R$ and
anti-neutrino luminosity $L_{\bar{\nu}}$ can be immediately
calculated, see Table~\ref{table:UThKinBSE} and, e.g.,
Ref.~\cite{Fiorentini:2004xk}.

\begin{table}[htb]
\caption{U, Th and K according to BSE, from
Ref.~\cite{Fiorentini:2004xk}.} \label{table:UThKinBSE}
\newcommand{\dg}{\hphantom{$0$}}
\newcommand{\cc}[1]{\multicolumn{1}{c}{#1}}
\renewcommand{\arraystretch}{1.2} 
\begin{tabular}{lccc}
\hline
       & \cc{$m$}            & \cc{$H_R$} & \cc{$L_{\nu}$} \\
       & \cc{[$10^{17}$~kg]} & \cc{[$10^{12}$~W]}
       & \cc{[$10^{24}$ s$^{-1}$]} \\
\hline
U        &  0.8 & 7.6 & \dg5.9 \\
Th       &  3.1 & 8.5 & \dg5.0 \\
$^{40}$K &  0.8 & 3.3 &   21.6 \\
\hline
\end{tabular}\\[2pt]
\end{table}

The BSE is a fundamental  geochemical paradigm. It is  consistent
with most  observations, which however regard  mostly the crust
and an undetermined portion of the mantle.  The measurement of
quantities --- such as the geo-neutrino signals --- which are
directly related to the global amounts of radioactive elements  in
the Earth  will provide a direct test of  this model for
composition and origin of  our planet.

For sure, heat released from radiogenic elements is a major source
of the terrestrial heat flow,  however its role  is not understood
at a quantitative level.  The masses estimated within the BSE
account for  the present radiogenic production of 19~TW, which is
about one half  of the estimated heat flow from
Earth~\cite{Anderson,Hofmeister}. Anderson refers to  this
difference as the missing heat source mystery and summarizes the
situation with the following words: ``Global heat flow estimates
range from 30 to 44~TW \ldots Estimates of the radiogenic
contribution \ldots based on cosmochemical considerations, vary
from 19 to 31~TW. Thus, there is either a good balance between
current input and output \ldots or there is a serious missing heat
source problem, up to a deficit of 25~TW \ldots'' If one can
determine   the amounts of radioactive elements by means of
geo-neutrinos, an important ingredient of Earth's energetics will
be fixed.

Concerning the distribution of radiogenic elements, estimates  for
Uranium in  the continental crust based on observational data are
in the range:
\begin{equation}\label{eq:masscontin}
    m_C=(0.3-0.4) \times 10^{17} \mathrm{\ kg} \quad .
\end{equation}
The extreme values have been obtained in
Ref.~\cite{Fiorentini:2004rj}  by taking the lowest (highest)
concentration reported in the literature for each layer of the
Earth's crust,  see Table~II of Ref.~\cite{Mantovani:2003yd}, and
integrating over a $2^\circ \times 2^\circ$ crust map.  The main
uncertainty  is from  the Uranium mass  abundance $a_{LC}$ in the
lower crust,  with estimates in the range $(0.2-1.1)$~ppm.
Estimates for the abundance in the upper crust, $a_{UC}$, are more
concordant, ranging from 2.2~ppm to 2.8~ppm. The crust --- really
a tiny envelope --- should thus contain  about one half of the BSE
prediction of Uranium in the Earth.

About the mantle, observational data are scarce and restricted to
the uppermost part, so the best estimate for its Uranium content
$m_M$ is obtained by subtracting the crust contribution from the
BSE estimate:
\begin{equation}\label{eq:massMantle}
    m_M = m_{\mathrm{BSE}} - m_C \quad .
\end{equation}

A commonly held view is that there is a vertical gradient in the
abundances of incompatible elements in the mantle, with the top
being most depleted. A minimum gradient model has a fully mixed
and globally homogeneous mantle; the other extreme is a model
where all the Uranium is at the bottom of the mantle.

Geochemical arguments are against the presence of radioactive
elements in the completely unexplored core, as discussed by
McDonough  in a recent review  of compositional models of the
Earth~\cite{Mcdonough:2003}.

Similar considerations hold for Thorium and Potassium,  the
relative mass abundance with respect to Uranium being globally
estimated as:
\begin{equation}\label{eq:ratioUThK}
    a(\mathrm{Th}):a(\mathrm{U}):a(\mathrm{K})\approx 4:1:10,000 \quad .
\end{equation}
We remark that the well-fixed ratios in Eq.~(\ref{eq:ratioUThK})
imply that detection of geo-neutrinos from Uranium will also bring
important information on the amount of Thorium and Potassium in
the whole Earth.

\begin{figure}[htbp]
\includegraphics[width=0.7\textwidth]{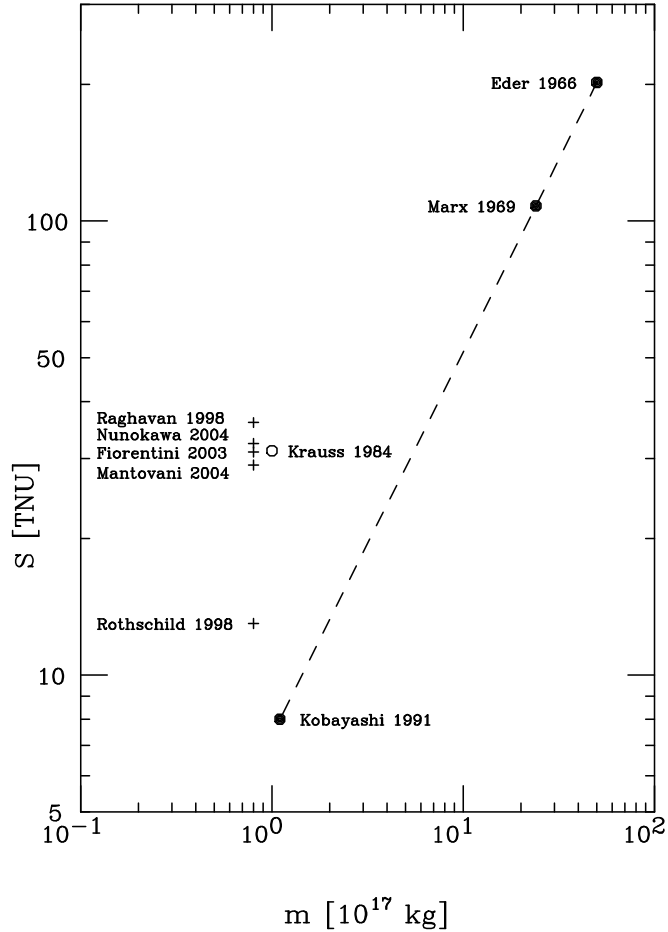}
 \caption{Previous estimates of the geo-neutrino signal $S$,
 renormalized to the average survival probability $\langle P_{ee}\rangle=0.59$, and
 the corresponding estimated Uranium mass $m$. The signal is in Terrestrial
 Neutrino Units (1 TNU = 1 event/year/10$^{32}$ proton).}
 \label{fig:previousEstimatesGeosignal}
\end{figure}

Several predictions for the geo-neutrino signal have been
presented in the past, corresponding to different hypotheses about
the amount of Uranium in the Earth and to different models of its
distribution. A summary is presented in
Fig.~\ref{fig:previousEstimatesGeosignal}. Early
models~\cite{Eder,Marx,Kobayashi} (full circles) assumed a uniform
Uranium distribution in the Earth and different values of the
Uranium mass. In fact these predictions are almost proportional to
$m$. The huge signals predicted by Eder and by Marx were obtained
by assuming that the Uranium density in the whole Earth is about
the same as that observed in the continental crust; Marx
(Eder~\footnote{The factor 10.7 appearing in the last of the
equations (13) of Ref.~\cite{Eder} should actually be 1.07 and,
correspondingly, the reaction rate on page~661 has been divided by
a factor of 10.}) assumed thus an Uranium mass 30 (60) times
larger than that estimated within the BSE.

Krauss et al.~\cite{Krauss:zn}  distributed about $10^{17}$~kg of
Uranium uniformly over a 30~km crust.   The other estimates
(crosses) are all obtained by using the BSE value for the mass as
an input and different models for distributing the Uranium content
between crust and mantle. In this class, Rotschild et
al.~\cite{Rothschild:1997dd}  obtained the minimal prediction by
assuming for the crust a very small Uranium abundance, definitely
lower  than  the values reported in more recent and detailed
estimates.

In this paper we shall use a rather  general approach, by keeping
the total  amount of Uranium  as a free variable, within  the
loose constraints provided from one side by the amount observed in
the crust and from the other side by the amount  tolerated by
Earth's energetics. We shall distribute the total amount between
crust and mantle so as to maximize  or minimize  the signal,
within the boundary provided by geochemical and/or geophysical
observations.

\section{\label{sec:nearTheDetector}The region near the detector}
As mentioned in the introduction,  the entire Earth's crust will
be subdivided into $2^\circ \times 2^\circ$  tiles. Within each
tile, one distinguishes several vertical layers and assigns to
each layer a world averaged Uranium mass abundance, see
Ref.~\cite{Mantovani:2003yd}. With the aim of reducing the error
on the regional contribution to the level of the uncertainty on
the rest of the world, one needs a more detailed geochemical and
geophysical study of the crust in the region within a few hundreds
kilometers from the detector, where some half of the signal is
generated.

We shall present here our results for the region near the KamLAND
detector, located at $36^\circ~25'~26"$~N and
$137^\circ~19'~11"$~E. We analyze the six tiles (see
Fig.~\ref{fig:ConradJapan}) around KamLAND by using geochemical
information on a  $1/4^\circ \times 1/4^\circ$ grid and a detailed
map of the crust depth. The possible (minimal and maximal) effects
of the subducting slab beneath Japan are also considered and the
uncertainty arising from the debated (continental or oceanic)
nature of the crust below the Japan Sea is taken into account.

\subsection{The six tiles near KamLAND}
\begin{figure}[htbpp]
\includegraphics[width=\textwidth,angle=0]{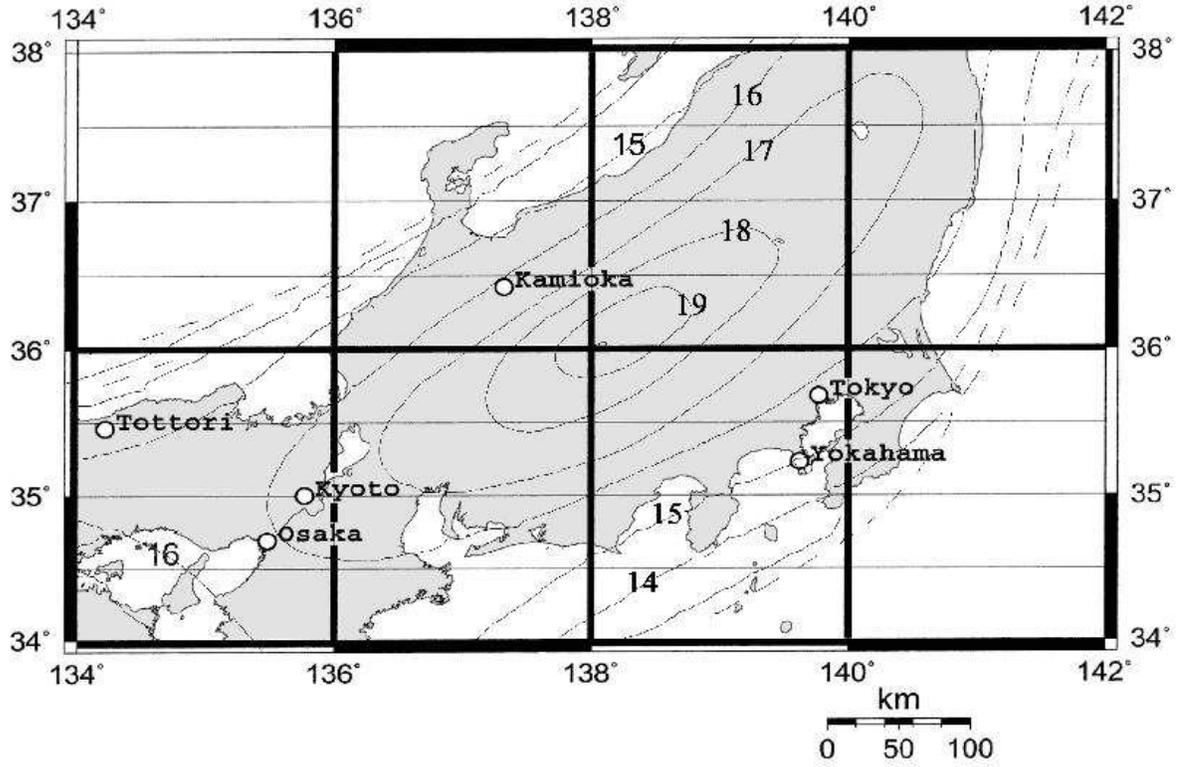}
 \caption{Depth of the Conrad discontinuity in Japan, from Ref.~\cite{Zhao:1992}.}
 \label{fig:ConradJapan}
\end{figure}

\begin{figure}[htbpp]
\includegraphics[width=\textwidth,angle=0]{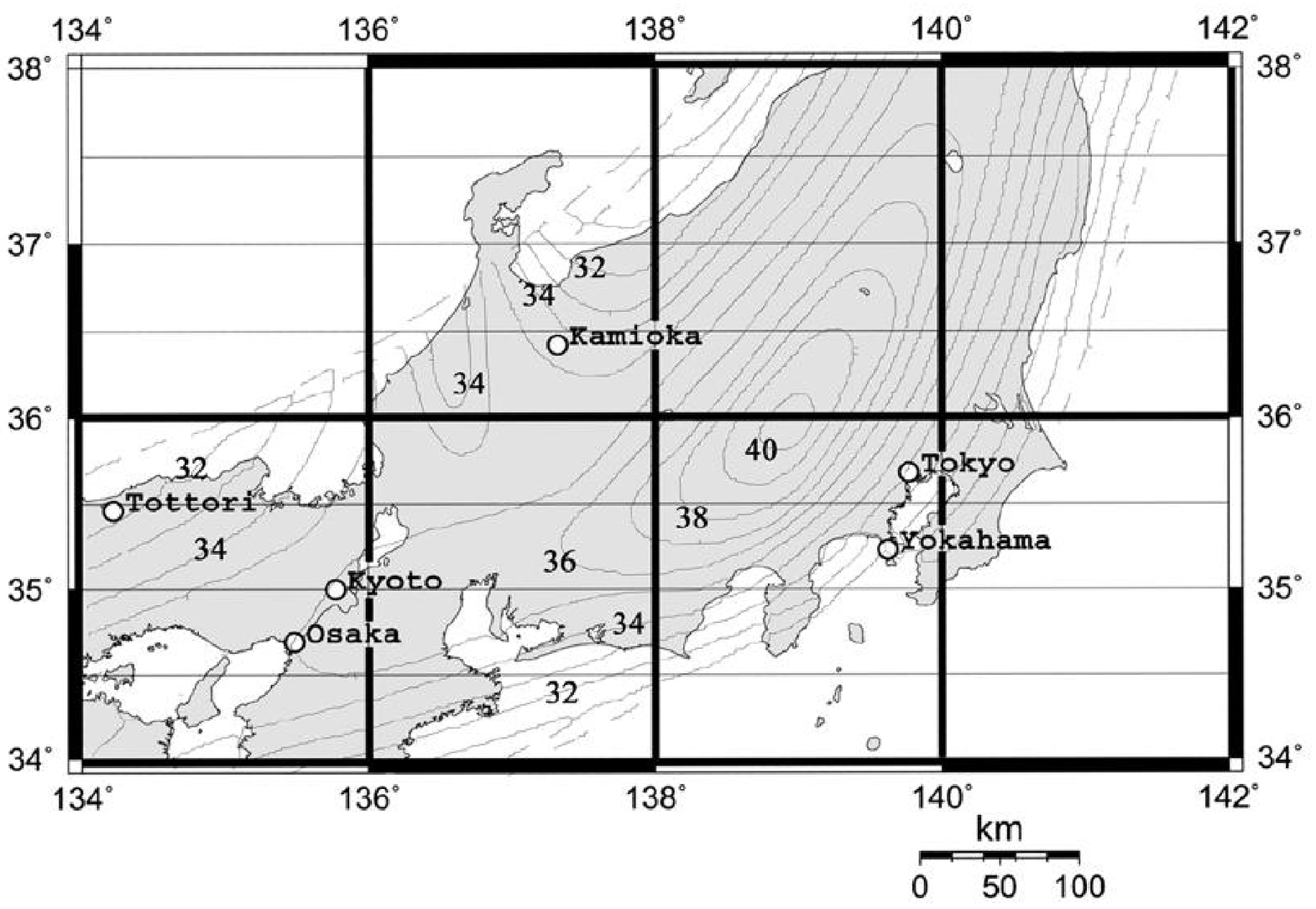}
\caption{Depth of the Moho discontinuity in Japan, from
Ref.~\cite{Zhao:1992}.}
 \label{fig:MohoJapan}
\end{figure}

The seismic velocity structure of the crust beneath the Japan
Islands has been determined  in Ref.~\cite{Zhao:1992} from the
study of some 13,000 arrival times of 562 local shallow
earthquakes. By applying an inversion method, the depth
distribution of the Conrad and Moho discontinuities beneath the
whole of the Japan Islands are derived, with an estimated standard
error  of $\pm 1$~km over most of Japan territory. Our
Figs.~\ref{fig:ConradJapan} and \ref{fig:MohoJapan}  are derived
from Fig.~6 of Ref.~\cite{Zhao:1992}. This allows distinguishing
two layers in the crust: an upper crust extending down to the
Conrad and a lower part down to the Moho discontinuity.

\begin{figure}[htbpp]
\includegraphics[width=\textwidth,angle=0]{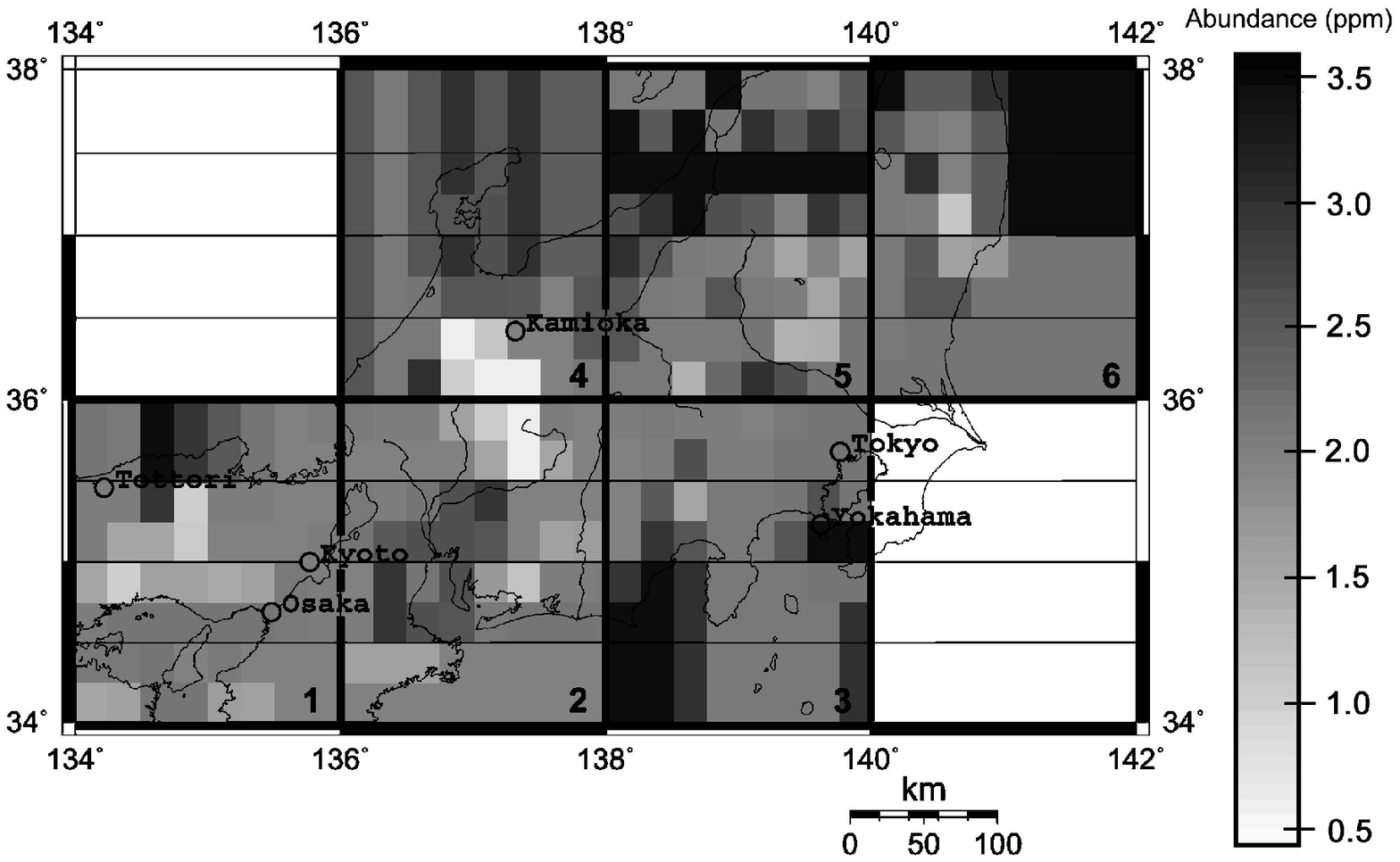}
\caption{Uranium abundance in the upper crust of Japan}
\label{fig:UraniumAbundanceJapan}
\end{figure}

The upper-crust  chemical composition  of Japan Islands has been
studied in Ref.~\cite{Togashi:2000}, based  on 166 representative
specimens, which can be associated  with 37 geological groups
based on ages, lithologies and provinces. By combining  the base
geological map of Fig.~2 of  Ref.~\cite{Togashi:2000} --- which
distinguishes 10 geological classes --- with the abundances
reported in Table~1 of the same paper, one can build a map of
Uranium abundance in the upper crust, under the \emph{important}
assumption that  the composition of the whole upper crust is the
same as that inferred in Ref.~\cite{Togashi:2000} from the study
of the exposed portion, see Fig.~\ref{fig:UraniumAbundanceJapan}.

We are not aware of a specific study  of the lower part of the
Japan crust, however,   it is well known that there are
similarities between the composition of  the Japanese crust and
that of the Sino-Korean block.   In an extensive compositional
study of East China crust~\cite{Gao:1998}, the  Uranium abundance
in the lower part is estimated between 0.63 and 1.08 ppm.  On
these grounds we shall take for the abundance  in the lower crust
of Japan:
\begin{equation}\label{eq:lowerCrustJapan}
    a_{LC}=0.85\pm 0.23 \quad .
\end{equation}

We remark that the estimated range of abundances for East China is
substantially narrower than the range for the whole world, which
is (0.2-1.1)~ppm (see Table~II in Ref.~\cite{Mantovani:2003yd}),
this last interval presumably reflecting regional differences in
the lower crust composition.

Concerning the vertical distribution of abundances in the crust,
it is presently impossible to have information  on   the chemical
composition on a scale smaller than the Conrad depth, generally
lying at about 20~km. We also note that the sampling density for
the study of the upper crust in the region near Kamioka is about
one specimen per 400~km$^2$.  On these grounds,    we introduce a
grid where one assigns a specific abundance to \emph{cells} with
size $1/4^\circ\times1/4^\circ$, i.e., with a linear scale of
about 20~km. Within each cell, the depth of the upper and lower
crust are taken from Ref.~\cite{Zhao:1992},  Uranium abundance for
the upper crust is derived from Ref.~\cite{Togashi:2000} and for
the lower crust from Ref.~\cite{Gao:1998}. In this way each of the
six tiles near Kamioka is subdivided  into sixty-four cells.

Just for the sake of computing  flux and signal,  each cell is
further subdivided into many \emph{subcells} with abundance
derived from those of the parent cell. The angle-integrated
produced  flux  at distance $R$ from a subcell of volume $\Delta
V$, calculated as that from a sphere with radius $r = (3/4 \pi
\Delta V)^{1/3}$, is
\begin{equation}\label{eq:fluxSubCell}
    \Delta\Phi = \frac{A}{4R}\left[ 2R r +
    (R^2-r^2)\ln\frac{|R-r|}{R+r}\right]
    \quad ,
\end{equation}
where $A$ is the specific activity (number of neutrinos produced
per unit time and volume).  Each subcell provides a contribution
to the signal rate  $\Delta S$ given by:
\begin{equation}\label{eq:signalSubCell}
    \Delta S = N_p \Delta\Phi \langle \sigma P_{ee}(R)\rangle \quad ,
\end{equation}
where  $N_p$  is the number of free protons in the target,
$\sigma$ is the cross section of reaction in
Eq.~(\ref{eq:inversebeta}), $P_{ee}(R)$ is the survival
probability which we shall calculate for $\tan^2\theta  = 0.40$
and $\Delta m^2=7.9\times 10^{-5}$~eV$^2$~\cite{Araki:2004mb}. The
average is over the energy spectrum of the neutrinos from Uranium
decay chain.

The resulting signal is obtained by adding  the contributions of
all subcells and will be expressed in Terrestrial Neutrino Units
(TNU), where 1~TNU corresponds to 1 event per year and per
$10^{32}$ protons. The contributions to the  produced flux and to
the signal from the six tiles are:
\begin{subequations}
\begin{eqnarray}
  \Phi_6 &=& 1.59 \times 10^{6} \mathrm{\ cm}^{-2}\mathrm{s}^{-1}\quad ; \\
  S_6    &=& 12.74 \mathrm{\ TNU} \quad .
\end{eqnarray}
\end{subequations}
With respect to our previous estimate from the whole
globe~\cite{Mantovani:2003yd}, giving $\Phi =3.676\times
10^{6}$~cm$^{-2}$~s$^{-1}$  and $S=28.6$~TNU for $\langle
P_{ee}\rangle = 0.59$, we find that the six tiles contribute 43\%
of the flux and 45\% of the signal: this justifies the close
scrutiny of the region within the six tiles. Some 3/4 of the
contribution arises form the upper crust.

In more detail, the tile hosting Kamioka  generates  29\% and 30\%
of the total produced flux and signal, respectively. The host
cell, i.e., the cell where Kamioka is located, contributes 9\% to
the total produced flux.

The  Uranium mass contained in the six tiles is about $m_6=
3.3\times 10^{13}$~kg, really  negligible (less then 0.05\%) with
respect to that estimated for the whole Earth.

We have considered several sources of the uncertainties  affecting
this estimate, see Table~\ref{table:nearDetectorRegion}:

\emph{(i) Measurement errors of the chemical analysis.} The
standard error of  the Uranium abundance measurements  in the
individual samples  is  3-4\%~\cite{Togashi:2000}. We translated
this error into a $\pm10\%$ global uncertainty (correspondingly to
about $3\sigma$)  on the Uranium abundance in the upper crust for
the six tiles~\footnote{This choice is very conservative given our
lack of information on the correlation between the errors of the
166 samples: the errors should partially average out and result in
a total error between 12\% and $(12/\sqrt{166})\%\approx 1\%$.}.

\emph{(ii) Discretization of the upper crust.} As mentioned above,
we divided the crust into $1/4^\circ \times 1/4^\circ$ cells,
assigning specific abundances to each cell. This discretization
procedure introduces some uncertainty, which is especially
important in the region very close to the detector. We have
evaluated the effect of replacing the abundance of the host cell
with those of adjacent cells. This produces signal variations in
the interval $(-0.64,+1.68)$~TNU. For simplicity, we introduce a
symmetrical error such as to encompass the extreme values.

\emph{(iii) Chemical composition of the  lower crust.} The error
is taken as the half-difference between signals obtained  for the
extreme values of the estimated Uranium abundances in the lower
crust.

\emph{(iv) Crustal depth.}  Since the depths of the Conrad and the
Moho discontinuities are estimated with an $3\sigma$ accuracy of
about $\pm3$~km, we have evaluated the effect of such (global)
variations over the six tiles, the error being again estimated as
the half difference between extreme values of flux/signal.

\begin{table}[htb] \caption{Errors from the regional geophysical
and geochemical uncertainties.} \label{table:nearDetectorRegion}
\newcommand{\dg}{\hphantom{$0$}}
\newcommand{\cc}[1]{\multicolumn{1}{c}{#1}}
\renewcommand{\arraystretch}{1.2} 
\begin{tabular}{lcl}
\hline
Source & $\Delta S$~(TNU) & Remarks \\
 \hline
 Composition of
upper-crust samples                    &  0.96   & $3\sigma$ error\\
Upper-crust discretizaion              &  1.68   &            \\
Lower-crust composition                &  0.82   & Full range \\
Crustal depths                         &  0.72   & $3\sigma$ error           \\
Subducting slab                        &  2.10   & Full range \\
Japan Sea                              &  0.31   & Full Range \\
\hline Total                           &  \bf{3.07}   &            \\
 \hline
\end{tabular}\\[2pt]
\end{table}

\subsection{The effect of the subducting slab beneath Japan}

The Japan arc, at the crossing   among the Eurasian, Philippine
and Pacific plates, is the theater of important subduction
processes.  The Philippine plate is moving towards the Eurasia
plate at about 40 mm/yr and is subducting beneath the southern
part of Japan. The Pacific Plate is moving in roughly the same
direction at about 80 mm/yr and is subducting beneath the northern
half of Japan.

We shall model these processes as a single slab penetrating below
Japan with velocity $v=60$~mm/yr, the average of the two plates.
This process has been occurring on a time scale $T\approx 10^8$~y,
during which the front has advanced by $D=vT\approx 6000$~km.

We assume that the slab brings with it   a sediments layer (with
density $\rho_{\mathrm{sed}}\approx 1.6$~ton/m$^{3}$, depth
$h_{\mathrm{sed}}\approx 350$~m and Uranium abundance
$a_{\mathrm{sed}}=1.4$~ppm, according to the data for the Japan
trench~\cite{Plank:1998}) on top of an oceanic crust layer, with
density $\rho_{OC}\approx 2.9$~ton/m$^{3}$,  vertical extension
$h_{OC}\approx 6.5$~km, and Uranium abundance $a_{OC}=0.1$~ppm.

The amount of Uranium carried by the slab per unit surface is
thus:
\begin{equation}\label{eq:slabUraniumMassPerUnit}
\sigma_{\mathrm{slab}} =a_{\mathrm{sed}}\times \rho_{\mathrm{sed}}
\times h_{\mathrm{sed}} + a_{OC} \times \rho_{OC} \times h_{OC} =
2.7 \mathrm{\ kg/m}^{2} \quad .
\end{equation}

We observe that the corresponding quantity for the lower
continental crust of  Japan (density $\rho_{LC}\approx
2.7$~ton/m$^{3}$, average depth $h_{LC}\approx 19$~km, and Uranium
abundance $a_{LC}=0.85$~ppm) is:
\begin{equation}\label{eq:slabLC}
\sigma_{LC} = a_{LC} \times \rho_{LC} \times h_{LC} = 43.6
\mathrm{\ kg/m}^{2} \quad .
\end{equation}

In order to estimate the effect of the subducting slab on
geo-neutrino production, one can envisage two extreme cases:

(a) one  assumes  that the slab keeps its trace elements while
subducting. The effect of its presence  can be estimated as if the
lower crust is effectively enriched by the amount of Uranium
contained in the crustal part of the  slab  passing below it,
i.e., $\sigma_{LC}\to \sigma_{LC} + \sigma_{\mathrm{slab}}$, a
negligible effect in comparison with the 25\% uncertainty on $
\sigma_{LC}$ resulting from $\Delta a_{LC}\approx 0.2$~ppm. The
signal is increased by 0.2~TNU.

(b) At the other extreme, it is possible that, as the slab
advances, \emph{all} Uranium from the subducting crust is
dissolved in fluids during dehydration reactions and accumulates
in the lower part of the continental crust of Japan, thus strongly
enriching it. The release process --- to a first approximation ---
will be uniform along the subduction direction, for some distance
$d\approx 250$~km corresponding to the dimension of  the Japan arc
transverse to the trench. Since  the slab has advanced by
$D\approx 6000$~km, the abundance increase in the lower crust is
now:
\begin{equation}\label{eq:slabExtreme2}
    \delta \sigma_{LC} = \sigma_{\mathrm{slab}}
    \times\frac{D}{d}\approx 68.4\mathrm{\ kg/m}^{2}    \quad .
\end{equation}
This corresponds to a substantial  increase of the effective
Uranium abundance in the Japanese lower continental crust.  The
signal contributed from the lower crust  of Japan increases by
4.4~TNU.

As we have no argument for deciding which of the extreme cases (a)
or (b) is closer to reality and  in order to encompass both of
them, we estimate the  contribution from the subducting slab as:
\begin{equation}\label{eq:slabEstimate}
    S_{\mathrm{slab}} = (2.3\pm 2.1)\mathrm{\ TNU} \quad .
\end{equation}

\subsection{The  crust below the Japan Sea}
The  morphology of the Japan  Sea,  characterized by three major
basins (Japan, Yamato, and Ulleung Basins) and topographic highs
such as the Yamato Ridge, is suggestive of intricate back-arc
opening tectonics.  Based on seismic reflection/refraction survey
data, bottom sampling data, geomagnetic data and basement depth
and topography, Tamaki et al.~\cite{Tamaki:1992} distinguish four
crustal types: continental, rifted continental, extended
continental,  and oceanic crust. The  Japan basin  is generally
considered as oceanic, whereas the nature of other basins is
controversial and debated.  Again, we resort  to two extreme
models:

(a) following Ref.~\cite{mappa} we consider all the basins as
formed with oceanic crust, extending down to 7~km below 1~km of
sediments. This provides a model for minimal geo-neutrino
production, resulting in $S_{JS}=0.06$~TNU.

(b) Deeper crustal depths (up to 19~km for the Oki bank) and
thicker sediments layers (up to 4~km for the Ulleung basin)  are
reported in the literature, see
Table~\ref{table:vertExtensCrustLayers}.  By taking these values
and assigning the abundances typical of continental crust, we
maximize geo-neutrino production, with the result
$S_{JS}=0.68$~TNU.

In between (a) and (b), and in order to encompass the extreme
values, we fix the contribution to the signal from the Japan Sea
as:
\begin{equation}\label{eq:JapanSeaEstimate}
    S_{JS} = (0.37\pm 0.31)\mathrm{\ TNU} \quad .
\end{equation}

\begin{table}[htb]
\caption{The vertical extensions  (km) of crustal layers in the
Yamano basin (YB), Oki bank (OK), and Ulleung basin (UB) used for
model (b).} \label{table:vertExtensCrustLayers}
\newcommand{\dg}{\hphantom{$0$}}
\newcommand{\cc}[1]{\multicolumn{1}{c}{#1}}
\renewcommand{\arraystretch}{1.2} 
\begin{tabular}{lccc}
\hline
       & YB            & OK & UB \\
\hline
sediments   &  1.2 & \dg0.3 & 4 \\
upper       &  2.8 & \dg8.7 & 2 \\
lower       &  8.5 &   10.5 & 8 \\
\hline
\end{tabular}\\[2pt]
\end{table}

\subsection{Summary of the regional contribution}

The regional contribution to the  signal  can be determined by
adding the previous results. As the  errors are independent, we
combine them in quadrature, obtaining:
\begin{equation}\label{eq:regionalContribution}
    S_{\mathrm{reg}} = (15.41\pm 3.07)\mathrm{\ TNU} \quad .
\end{equation}
The principal uncertainty comes from the effect of the subducting
slab. A more detailed study of the mechanisms of Uranium release
should exclude the extreme cases which we have considered, thus
reducing the error.

Discretization of the crust is major uncertainty. A more detailed
description of the exposed crust is certainly achievable, however,
it will bring little help without a better understanding of the
chemical composition  variation with depth.

\section{\label{sec:restOfTheWorld}The rest of the world}
The contribution from the rest of the world will depend on  the
total mass of Uranium $m$ as well as on its distribution inside
the Earth, since the closer is the source to the detector the
larger is its contribution to the signal. For each value of $m$,
we shall construct the distributions which provide  the maximal
and minimal signals  under the condition that they are consistent
with geochemical and geophysical information on the globe.

For the Earth's crust, we use the  $2^\circ \times 2^\circ$ map of
Ref.~\cite{mappa} distinguishing several crustal layers which are
known to contain different amounts of radioactive elements. For
each layer minimal and maximal estimates of Uranium abundances
found in the literature are adopted, so as to obtain a range of
acceptable fluxes,  see Table~\ref{table:minMaxU}.

\begin{table}[htb]
\caption{Minimal and maximal estimated Uranium abundances for the
continental crust, in ppm.} \label{table:minMaxU}
\newcommand{\dg}{\hphantom{$0$}}
\newcommand{\cc}[1]{\multicolumn{1}{c}{#1}}
\renewcommand{\arraystretch}{1.2} 
\begin{tabular}{lcc}
\hline
       & min            & max \\
\hline
upper crust     &  2.2 & 2.8 \\
lower crust     &  0.2 & 1.1 \\
\hline
\end{tabular}\\[2pt]
\end{table}

Depending on the adopted values, the Uranium mass in the crust
$m_C$ is the range  $(0.3-0.4)$ in units --- here and in the
following --- of $10^{17}$~kg. Clearly the larger is the mass the
bigger is the signal, the extreme values being~\footnote{As we are
now considering distances from the detector which are considerably
larger than the neutrino oscillation length, the asymptotic
expression for the survival probability, $\langle P_{ee}\rangle =
(1-1/2 \sin^2(2\theta))=(1+\tan^4\theta)/(1+\tan^2\theta)^2$
holds, so that the produced flux and signal are directly
proportional, $S/\mathrm{TNU} = 13.2 \times \langle P_{ee}\rangle
\Phi / (10^6 \mathrm{cm}^{-2}
\mathrm{s}^{-1})$~\cite{Fiorentini:2003ww}.}:
\begin{equation}\label{eq:segnaleCrostaMinMax}
S_C^{\mathrm{(min)}} = 6.448\quad\mathrm{for\ }m=0.3 \quad\quad
\mathrm{and} \quad\quad S_C^{\mathrm{(max)}} =
8.652\quad\mathrm{for\ }m=0.4 \, .
\end{equation}

Concerning Uranium in the mantle,  we  assume that spherical
symmetry holds  and that the Uranium mass abundance is a
non-decreasing function of depth. It follows that, for  a fixed
Uranium mass in the mantle  $m_M$,  the extreme predictions for
the signal are obtained by: (i) placing Uranium in a thin layer at
the bottom and (ii) distributing it with uniform abundance over
the mantle. These two cases give, respectively:
\begin{equation}\label{eq:segnaleMantelloMinMax}
S_M^{\mathrm{(min)}} = 12.15 \times m_M\mathrm{\ TNU} \quad\quad
\mathrm{and} \quad\quad S_M^{\mathrm{(max)}} = 17.37 \times
m_M\mathrm{\ TNU} \, .
\end{equation}

By using again the proximity  argument,  we can  combine the
contributions  from crust and mantle so as to obtain extreme
predictions: for a fixed total $m= m_C+m_M$, the highest signal is
obtained by assigning to the crust as much material  as consistent
with observational data ($m_C=0.4$) and putting  the rest,
$m-m_C$, in the mantle with a uniform distribution. Similarly, the
minimal flux/signal is obtained for the minimal mass in the crust
($m_C= 0.3$) and the rest in a thin layer at the bottom of the
mantle. In conclusion, the contribution from the rest of the world
is in the range:

\begin{equation}\label{eq:segnaleRestOfTheWorldMinMax}
S_{RW}^{\mathrm{(min)}} = \left[6.448+12.15(m-0.3)\right]\mathrm{\
TNU}
 \quad
\mathrm{and} \quad S_{RW}^{\mathrm{(max)}} =
\left[8.652+17.37(m-0.4)\right]\mathrm{\ TNU} \, .
\end{equation}

Note that the two straight lines cross near $m=0.21$, i.e., in the
nonphysical region, since Uranium mass is at least 0.3.

 We remind that  that both  the Uranium poor ($m_C=0.3$) and the
Uranium rich ($m_C=0.4$) crust models are observationally
acceptable.   We also  recall  that a compositionally uniform
mantle is advocated  by geophysicists, whereas geochemists  prefer
a two-layered mantle with a lower part close to the primitive
composition   and an  upper part strongly impoverished in Uranium.
In other words, it seems to us that the extreme predictions
correspond to equally plausible models.    On these  grounds, we
take as our prediction  the mean of the extremes  and assign an
error so as to encompass both of them:

\begin{eqnarray}\label{eq:signalRestOfWorld}
S_{RW} =  \left[ (2.25 + 14.76\times m) \pm ( -0.55 + 2.61\times
m)\right]\mathrm{\ TNU} \quad .
\end{eqnarray}

We remark that by combining global mass balance with geometry, we
have strongly  constrained  the contribution from the rest of the
world: for a mass near the BSE values, $m\approx 0.8$, the signal
is predicted within about 10\%.

\section{\label{sec:signalVsMass}
The geo-neutrino signal as a function of Uranium mass in the
Earth}

By combining  the regional contribution,
Eq.~(\ref{eq:regionalContribution}), with that from the rest of
the world, Eq.~(\ref{eq:signalRestOfWorld}), we get the Uranium
geo-neutrino signal as a function of Uranium mass in the Earth:
\begin{subequations}
\begin{equation}\label{eq:signalTotal}
    S=S_0 \pm \Delta \quad ,
\end{equation}
 where:
\begin{eqnarray}
  S_0    &=&  17.66 + 14.76 \times m
\label{eq:totalGeoSignal}\\
  \Delta^2 &=& (3.07)^2 +(2.61\times m - 0.55)^2
 \quad . \label{eq:DeltaTotalGeoSignal}
\end{eqnarray}
\end{subequations}

 This error is obtained by combining in quadrature  all
geochemical and geophysical uncertainties discussed in the
preceding paragraphs. All of them have been estimated so as to
cover $\pm3\sigma$ intervals of experimental measurements and
total ranges of theoretical predictions.

However, this error does not account for present uncertainties on
neutrino oscillation parameters and on the cross section of
reaction in Eq.~(\ref{eq:inversebeta}). For the sake of discussing
the potential of geo-neutrinos, we shall ignore for the moment
these error sources.

The expected signal from Uranium geo-neutrinos at  KamLAND   is
presented as a function of the total Uranium mass $m$ in
Fig.~\ref{fig:Kamlandsignalmass}.  The upper horizontal scale
indicates the corresponding radiogenic heat production rate from
Uranium ($H_R=9.5\times m$).

The predicted signal as a function of $m$  is between the two
lines denoted  as $S_{\mathrm{high}}$ and $S_{\mathrm{low}}$,
which correspond, respectively, to $S_0 \pm \Delta$.

Since the minimal amount of Uranium in the Earth is $0.3\times
10^{17}$~kg (corresponding to the minimal estimate for the crust
and the assumption of  negligible amount in the mantle), we expect
a signal of at least 19~TNU. On the other hand, the maximal amount
of Uranium tolerated by Earth's energetics~\footnote{For an
Uranium mass  $m=1.8\times 10^{17}$~kg and relative abundances as
in Eq.~(\ref{eq:ratioUThK}), the present radiogenic heat
production rate from U, Th and K decays equals the maximal
estimate for the present heat flow from Earth,
$H_E^{\mathrm{max}}=44$~TW~\cite{Pollack:1993}.}, $1.8\times
10^{17}$~kg, implies a signal not exceeding 49~TNU.

For the central value of the BSE model, $m=0.8\times 10^{17}$~kg,
we predict $S=29.5\pm 3.4$~TNU, i.e., with  an accuracy of 12\% at
``$3\sigma$''. We remark that estimates by different authors for
the Uranium mass within the BSE are all between $(0.7-0.9)\times
10^{17}$~kg. This implies that the   Uranium signal has to be in
the interval $(24.7-34.5)$~TNU. The measurement of geo-neutrinos
can  thus provide a direct test of an important geochemical
paradigm.

The effect of uncertainties about the oscillation parameters is
presented in Table~\ref{table:oscillationParameters}. In this
respect the mixing angle is most important. Figure~4~(b) of
Ref.~\cite{Araki:2004mb} shows a $3\sigma$ range
$0.26<\tan^2\theta<0.67$ (central value 0.40): the corresponding
range for the average survival probability is $0.52<P_{ee}<0.67$
(central value 0.59), with a $3\sigma$ relative error on the
signal $\Delta S/S \approx 13\%$, which is comparable to the
geological uncertainty in Eq.~(\ref{eq:DeltaTotalGeoSignal}). The
mixing angle should be determined more precisely for fully
exploiting the geo-neutrino signal.

\begin{table}[htb] \caption{
Effect of the oscillation parameters on the signal. The
relative/absolute variation is computed with respect to the
prediction for the best fit values  ($\delta m^2 = 7.9\times
10^{-5}$~eV$^2$ and $\tan^2\theta = 0.40$). }
\label{table:oscillationParameters}
\newcommand{\dg}{\hphantom{$0$}}
\newcommand{\cc}[1]{\multicolumn{1}{c}{#1}}
\renewcommand{\arraystretch}{1.2} 
\begin{tabular}{lc}
 \hline\hline
parameter & signal variation \\
 \hline
$\tan^2\theta = 0.26$                  & $+13.5\%$ \\
$\tan^2\theta = 0.67$                  & $-12.2\%$ \\
\hline
$\delta m^2 = 6.9\times 10^{-5}$~eV$^2$ &  $+0.11$~TNU   \\
$\delta m^2 = 9.3\times 10^{-5}$~eV$^2$ &  $-0.10$~TNU   \\
\hline \hline
\end{tabular}\\[2pt]
\end{table}

On the other hand, the predicted signal is practically unaffected
by the uncertainty  on the neutrinos squared mass difference
$\delta m^2$: when this is varied  within its $\pm3\sigma$
interval the signal changes by 0.1~TNU. This holds  for any value
of the total Uranium mass $m$, since the precise value of $\delta
m^2$ only matters in the region near the detector. In addition, we
observe that the predictions computed for the best value ($\delta
m^2=7.9\times 10^{-5}$~eV$^2$) and for the limit $\delta
m^2=\infty$ differ by +0.3~TNU. Finally, the error on the
inverse-beta cross section (quoted as 0.2\% at $1\sigma$ in
Ref.~\cite{Eguchi:2002dm}) translates into a $3\sigma$ uncertainty
of 0.6\% on the signal.
\section{Concluding remarks}
We summarize here the main points of this paper:

(1) Based on a detailed geochemical and geophysical study of the
region near Kamioka, we have determined the regional contribution
to the signal from Uranium-geoneutrinos:
\begin{equation}\label{eq:regionalContributionConcl}
    S_{\mathrm{reg}} = (15.41\pm 2.98)\mathrm{\ TNU} \quad .
\end{equation}

(2) By using  global mass balance arguments, we have determined
the contribution from the rest of the world:
\begin{eqnarray}\label{eq:signalRestOfWorldConcl}
S_{RW} =  \left[ (2.25 + 14.76\times m) \pm ( -0.55 + 2.61\times
m)\right]\mathrm{\ TNU} \quad ,
\end{eqnarray}
where $m$ is the Uranium mass in the Earth, in units of
$10^{17}$~kg.

(3) Our prediction for the signal as a function of $m$ is
presented in Fig.~\ref{fig:Kamlandsignalmass}, which  shows  the
potential of geo-neutrinos for determining how much Uranium is in
the Earth (As discussed in the paper, the range of the prediction
for a given mass is mostly due to experimental determinations of
local abundances and to the geometrical distributions of trace
elements in the mantle).

(4) Measurements of the antineutrino signal from Uranium can
provide crucial tests for models. In particular, estimates by
different authors for the Uranium mass within the important
paradigm of the Bulk Silicate Earth are all in the range $(0.7,
0.9)\times 10^{17}$kg, which translates into a signal $23 <
S(\mathrm{U}) < 31$~TNU.

(5) A full exploitation of the geo-neutrino signal demands that
the mixing angle is determined more precisely.

\section*{Acknowledgments}
We are particularly grateful to Enomoto Sanshiro for useful
discussions and for pointing to our attention
Refs.~\cite{Zhao:1992} and \cite{Togashi:2000}. We thank
M.~Contorti, K.~Inoue, and E.~Lisi for valuable comments. The
final manuscript includes several useful remarks by the anonymous
referees.

This work was partially supported by MIUR (Ministero
dell'Istruzione, dell'Universit\`a e della Ricerca) under
MIUR-PRIN-2003 project ``Theoretical Physics of the Nucleus and
the Many-Body Systems'' and MIUR-PRIN-2004 project ``Astroparticle
physics''.

\section*{Note added in Proofs}
The KamLAND collaboration has just presented experimental
results~\cite{Araki:2005qa} on geo-neutrinos, which are in
agreement with our predictions, see Ref.~\cite{Fiorentini:2005ma}
for  a comparison.


\begin{thebibliography}{00}

\bibitem{Eder}
G.~Eder, Nucl. Phys. {\bf 78}, 657 (1966).

\bibitem{Marx}
G.~Marx, Czech. J. Phys. {\bf B 19}, 1471-1479 (1969).

\bibitem{Krauss:zn}
L.~M.~Krauss, S.~L.~Glashow, and D.~N.~Schramm,
Nature (London) {\bf 310}, 191 (1984).

\bibitem{Kobayashi}
M.~Kobayashi and Y.~Fukao, Geophysical Research Lett. {\bf 18},
633-636 (1991).

\bibitem{Raghavan:1997gw}
R.~S.~Raghavan, S.~Sch{\"o}nert, S.~Enomoto, J.~Shirai,
F.~Suekane, and A.~Suzuki,
Phys.\ Rev.\ Lett.\  {\bf 80}, 635 (1998).

\bibitem{Rothschild:1997dd}
C.~G.~Rothschild, M.~C.~Chen, and F.~P.~Calaprice,
Geophy. Res. Lett. {\bf 25}, 1083 (1998) [arXiv:nucl-ex/9710001].

\bibitem{Fiorentini:2002bp}
G.~Fiorentini, F.~Mantovani, and B.~Ricci,
Phys.\ Lett.\ B {\bf 557}, 139 (2003) [arXiv:nucl-ex/0212008].

\bibitem{Fiorentini:2003ww}
G.~Fiorentini, T.~Lasserre, M.~Lissia, B.~Ricci, and
S.~Sch{\"o}nert,
Phys.\ Lett.\ B {\bf 558}, 15 (2003) [arXiv:hep-ph/0301042].

\bibitem{Mantovani:2003yd}
F.~Mantovani, L.~Carmignani, G.~Fiorentini and M.~Lissia,
Phys.\ Rev.\ D {\bf 69}, 013001 (2004) [arXiv:hep-ph/0309013].


\bibitem{Fiorentini:2004xk}
G.~Fiorentini, M.~Lissia, F.~Mantovani and R.~Vannucci,
arXiv:hep-ph/0409152.

\bibitem{Fiorentini:2004rj}
G.~Fiorentini, M.~Lissia, F.~Mantovani and R.~Vannucci,
Astroparticle and high energy physics AHEP2003/035
[arXiv:hep-ph/0401085].

\bibitem{Eguchi:2002dm}
KamLAND Collaboration, K.~Eguchi {\it et al.},
Phys.\ Rev.\ Lett.\  {\bf 90}, 021802 (2003)
[arXiv:hep-ex/0212021].

\bibitem{Nunokawa:2003dd}
H.~Nunokawa, W.~J.~C.~Teves and R.~Zukanovich Funchal,
JHEP {\bf 0311}, 020 (2003) [arXiv:hep-ph/0308175].

\bibitem{Mitsui:2003fm}
T.~Mitsui  [KamLAND Collaboration],
{\it Proceedings of the 28th International Cosmic Ray Conferences
(ICRC 2003), Tsukuba, Japan, 31 Jul - 7 Aug 2003}, pp.~1221-1224.

\bibitem{Miramonti:2003hw}
L.~Miramonti,
arXiv:hep-ex/0307029.

\bibitem{Domogatski:2004gs}
G.~Domogatski, V.~Kopeikin, L.~Mikaelyan and V.~Sinev,
arXiv:hep-ph/0401221.

\bibitem{McKeown:2004yq}
R.~D.~McKeown and P.~Vogel,
Phys.\ Rept.\  {\bf 394}, 315 (2004) [arXiv:hep-ph/0402025].

\bibitem{Fields:2004tf}
B.~D.~Fields and K.~A.~Hochmuth,
arXiv:hep-ph/0406001.

\bibitem{Rusov:2003sx}
V.~D.~Rusov, V.~N.~Pavlovich, V.~N.~Vaschenko, V.~A.~Tarasov,
D.~A.~Litvinov, V.~N.~Bolshakov and E.~N.~Khotyaintseva,
arXiv:hep-ph/0312296.

\bibitem{Fogli:2004vb}
G.~L.~Fogli, E.~Lisi, A.~Palazzo and A.~M.~Rotunno,
arXiv:hep-ph/0405139.


\bibitem{Domogatski:2004zz}
G.~Domogatski, V.~Kopeikin, L.~Mikaelyan and V.~Sinev,
arXiv:hep-ph/0409069.


\bibitem{Pollack:1993}
H.~N.~Pollack, S.~J.~Hunter and J.~R.~Johnson,
Rev.\ Geophys.\ {\bf 31}, 267-280 (1993).

\bibitem{Anderson}
D.~Anderson, Energetics of the Earth and the Missing Heat Source
Mystery, available at www.mantleplumes.org/Energetics.html

\bibitem{Hofmeister}
A.~M.~Hofmeister and R.~E.Criss,
Tectonophysics {\bf 395}, 159--177 (2005).

\bibitem{mappa}
C.~Bassin,  G.~Laske. and G.~Masters,
EOS Trans. Am. Geophys. Union  {\bf 81}, F897  (2000)
[http://mahi.ucsd.edu/Gabi/rem.html].


\bibitem{Mcdonough:2003}
W.~F.~McDonough, Compositional Model for The Earth's Core,
pp.~547-568.  In The Mantle and Core (ed. R.W. Carlson.) Vol. 2
Treatise on Geochemistry (eds. H.D. Holland and K.K. Turekian),
Elsevier-Pergamon, Oxford (2003).


\bibitem{Zhao:1992}
D.~Zhao, S.~Horiuchi and A.~Hasegawa,
Tectonophysics {\bf 212}, 289-301 (1992).

\bibitem{Togashi:2000}
S.~Togashi, et al.,
Geochem. Geophys. Geosyst. (Electronic Journal of the Earth
Sciences) {\bf 1}, 2000GC00083 (2000).

\bibitem{Gao:1998}
S.~Gao, et al.,
Geochimica et Cosmochimica Acta {\bf 62}, 1959-1975 (1998).

\bibitem{Araki:2004mb}
T.~Araki {\it et al.}  [KamLAND Collaboration],
arXiv:hep-ex/0406035.

\bibitem{Plank:1998}
T.~Plank and C.~H.~Langmuir,
Chemical geology {\bf 145}, 325-394 (1998).

\bibitem{Tamaki:1992}
K.~Tamaki, K.~Suyehiro, J.~Allan, J.`C.~Ingle, and
K.~A.~Pisciotto,
in Proc. ODP, Sci. Results, edited by K.~Tamaki,  K.~Suyehiro,
J.~Allan, M.~McWilliams, et al., {\bf 127/128}, 1333-1348, College
Station, TX (Ocean Drilling TAMU), (1992).

\bibitem{Araki:2005qa}
T.~Araki {\it et al.}  [KamLAND Collaboration],
Nature {\bf 436}, 499 (2005).

\bibitem{Fiorentini:2005ma}
  G.~Fiorentini, M.~Lissia, F.~Mantovani and B.~Ricci,
  arXiv:hep-ph/0508048.

\end{thebibliography}
\end{document}